\documentclass{article}
\usepackage{spconf,amsmath,graphicx}

\usepackage{amsfonts} % mathbb
\usepackage{gensymb}  % for \degree
\usepackage{multicol}
\usepackage{multirow}

\usepackage{amssymb}% http://ctan.org/pkg/amssymb
\usepackage{pifont}% http://ctan.org/pkg/pifont

\newcommand\blfootnote[1]{%
  \begingroup
  \renewcommand\thefootnote{}\footnote{#1}%
  \addtocounter{footnote}{-1}%
  \endgroup
}
\usepackage{xpatch,xcolor}
\usepackage{hyperref}
\hypersetup{
    colorlinks=true,
    linkcolor=purple,
    filecolor=magenta,      
    urlcolor=purple,
}
\usepackage{bbm}

\usepackage{booktabs}
\usepackage{makecell}

\title{Neural Fourier Shift for Binaural Speech Rendering}

\name{Jin Woo Lee$^1$, Kyogu Lee$^{1,2,3}$}
\address{
    $^1$Department of Intelligence and Information, Seoul National University \\
    $^2$Interdisciplinary Program in Artificial Intelligence, Seoul National University \\
    $^3$AI Institute, Seoul National University \\
    {\tt\footnotesize\small\{jinwlee,kglee\}@snu.ac.kr}
}

\begin{document}
\ninept

\maketitle

\begin{abstract}
We present a neural network for rendering binaural speech from given monaural audio, position, and orientation of the source. Most of the previous works have focused on synthesizing binaural speeches by conditioning the positions and orientations in the feature space of convolutional neural networks. These synthesis approaches are powerful in estimating the target binaural speeches even for in-the-wild data but are difficult to generalize for rendering the audio from out-of-distribution domains. To alleviate this, we propose Neural Fourier Shift (NFS), a novel network architecture that enables binaural speech rendering in the Fourier space. Specifically, utilizing a geometric time delay based on the distance between the source and the receiver, NFS is trained to predict the delays and scales of various early reflections. NFS is efficient in both memory and computational cost, is interpretable, and operates independently of the source domain by its design. Experimental results show that NFS performs comparable to the previous studies on the benchmark dataset, even with its 25 times lighter memory and 6 times fewer calculations.
\end{abstract}
% Further experiments with NFS for binauralizing the out-of-distribution dataset including singing, music, and general noise data show that the proposed method also generalizes well.
%

\blfootnote{Demo page with sound samples: {\scriptsize\url{https://bit.ly/3Sy6iFa}}}

\vspace{-5mm}
\section{Introduction}

% Binaural audio synthesis
Binaural audio synthesis is an essential technology for providing immersive sound experiences in virtual and extended reality.
By simulating the precise clues to locate the sound source, binaural sound synthesis aims to provide a spatial experience to our auditory cues \cite{hendrix1996sense}, which is also known as spatial hearing.
Rendering spatialized audio from arbitrary sounds with low computational cost and high quality is an essential technology in AR/VR platforms.

% DSP -> neural
% DSP: LTI라서 안좋다는 의견이 제시되기도 한다.
Humans can localize sounds using a variety of perceptual cues, including Interaural Time Difference (ITD), Interaural Intensity Difference (IID), and spectral distortions caused by interactions with the listener's pinnae, head, and torso \cite{rayleigh1907xii,wright1974pinna,asano1990role}.
Such cues can implicitly be modeled using head-related impulse response (HRIR) and binaural room impulse response (BRIR) \cite{wightman1989headphone,moller1992fundamentals}.
Traditional approaches for binaural audio rendering relied on digital signal processing (DSP) theories to model the HRIRs, BRIRs, or ambient noises \cite{savioja1999creating,jianjun2015natural,zotkin2004rendering,lee2022global}.
Most of these approaches model the binaural environments based on the linear time-invariant (LTI) system, and some difficulties have been discussed that such methods are insufficient for accurately modeling the real-world acoustic nonlinearities \cite{brinkmann2017authenticity}.

% 이를 지적하며 NN-approach 등장
While pointing out such limitations, recent studies utilize neural networks on binaural speech synthesis \cite{richard2020neural,leng2022binauralgrad}.
By conditioning the positions and orientations in the feature space of convolutional neural networks, the neural network approaches synthesize the binaural speeches as raw waveforms.
These synthesis approaches are powerful in estimating the distribution of target real-world binaural speeches, but it is difficult to generalize for rendering the audio from out-of-distribution domains such as music, or general sound events.

\begin{figure}[t]
    \begin{minipage}[b]{1.0\linewidth}
        \centering
        \centerline{\includegraphics[width=0.75\linewidth]{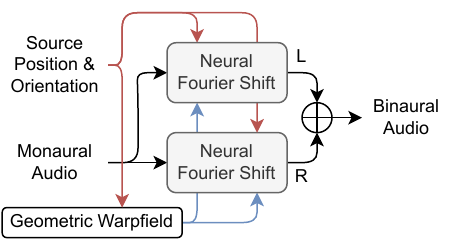}}
        \centerline{(a) Proposed system for binaural speech rendering}
        \medskip
    \end{minipage}
    \begin{minipage}[b]{1.0\linewidth}
        \centering
        \centerline{\includegraphics[width=0.65\linewidth]{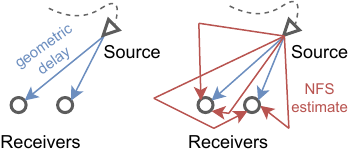}}
        \centerline{(b) Visualization of the proposed idea}
    \end{minipage}
    \caption{Schematic diagram of the proposed system and visualization of the modeled acoustic paths.
    % Utilizing geometric delays that reflect only the shortest path between the source and receiver, our model predicts other possible early reflections.
    }
    \label{fig:system}
\end{figure}

% Our work
In this paper, we propose a novel neural network that can render binaural speech independent of the source domain, based on the theoretical foundation of DSP.
We start from the idea to use the delays that can be computed from geometric distances as bases for early reflections.
% Such delay can be considered equivalent to the `Geometric Warpfield' of \cite{richard2020neural}, which represents the phase lag in the acoustic path arriving at the shortest of many possible paths between the source and receiver.
Such delay can be considered equivalent to the phase lag in the acoustic path arriving at the shortest of many possible paths between the source and receiver.
% Based on the computed shortest delay, we aim to predict the possible early reflections and model the responses.
We aim to predict the responses of the other acoustic paths such as early reflections.
Figure \ref{fig:system} shows a visualization of the exemplar acoustic paths.
To this end, we summarize our contributions as follows:
\begin{itemize}
    \item We propose a model for binaural rendering, which is highly efficient in memory and computational cost.
    %\item We empirically show that our method outperforms the previous studies in quantitative measures.
    \item We empirically show that our method achieves similar performance to prior studies at a much lower computational cost through quantitative and perceptual evaluation.
    %\item Perceptual evaluation shows the effect of the proposed architecture, along with the performance comparison.
\end{itemize}

\section{Related Works}

Traditionally, signal processing technology has long laid the foundation for binaural audio synthesis \cite{savioja1999creating,jianjun2015natural,zotkin2004rendering}.
However, more recent studies utilize deep learning to solve binaural speech synthesis in an end-to-end manner \cite{richard2020neural,leng2022binauralgrad,huang2022end}.
Although these data-driven approaches are powerful in estimating the target binaural speech, even for the in-the-wild binaural recordings, they are inherently limited to estimating the distribution of the training set.
Consequently, they are prone to failure in generalizations and may require other models to render out-of-domain data, which can be a disadvantage when considering on-device inference scenarios.

This situation naturally leads to the parameter estimation method as a way to utilize the powerful estimation ability of the neural network without being limited to the data domain.
Lee \textit{et al.} \cite{lee2022differentiable} showed that artificial reverberation parameters can be estimated in an end-to-end manner.
They leveraged neural networks to estimate the reverberation parameters analysis-synthesis task and outperformed the non-end-to-end approaches.
Focusing on the spatial audio, Gebru \textit{et al.} \cite{gebru2021implicit} implicitly approximated HRTFs using a neural network.
More recent work by Richard \textit{et al.} \cite{richard2022deep} proposed a framework for training neural networks to estimate IRs for the given position.
Their proposed IR-MLP efficiently estimate the IRs and even learns the IRs from an in-the-wild binaural dataset, but there is room for performance improvement as the authors discussed in their paper.

Fourier transforms have historically had a significant impact on engineering advances and are often used to improve the performance of deep learning techniques.
Fourier space is used for proving that the neural networks are universal approximators \cite{hornik1989multilayer}, and the transform advantages efficient computation of the neural convolutional layers \cite{mathieu2013fast}.
%More recently, it has also been reported that neural network architectures using Fourier transforms or using sinusoidal activation functions are effective in solving kernel regression or inverse problems \cite{tancik2020fourier,sitzmann2020implicit}.
More recently, Fourier transforms and sinusoidal activations are reported to benefit neural networks in solving kernel regression or inverse problems \cite{tancik2020fourier,sitzmann2020implicit}.
Neural operators also benefit from the capability to learn parametric dependence to the solution of partial differential equations in quasi-linear time complexity, when defined in Fourier space \cite{li2020fourier}.

\begin{figure}[t]
    \begin{minipage}[b]{1.0\linewidth}
        \centering
        \centerline{\includegraphics[width=0.9\linewidth]{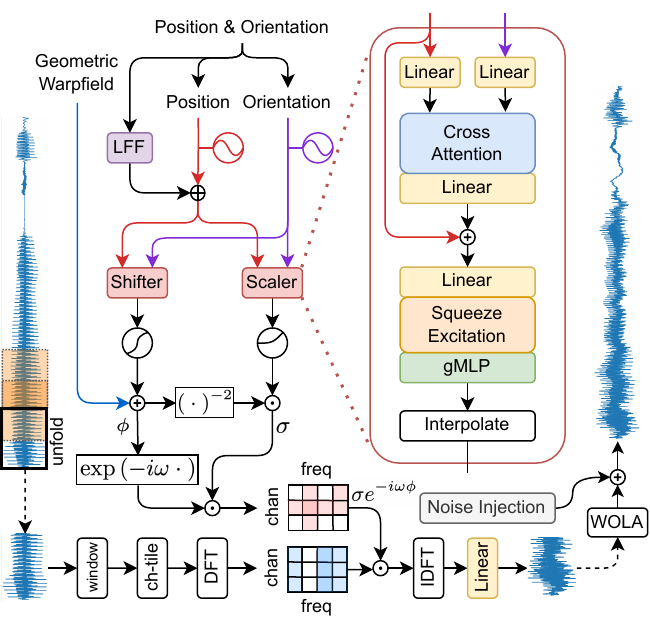}}
    \end{minipage}
    \vspace{-7mm}
    \caption{Architecture of NFS.
    Dashed arrows represent framewise segmentation or collection.
    {\tt\footnotesize ch-tile} indicates channel-wise repetition.
    Every {\tt\footnotesize Linear} indicates a channel-wise linear operation.
    }
    \label{fig:nfs}
    \vspace{-4mm}
\end{figure}

\section{Proposed Method}

This work mainly focuses on modeling binaural speech using a neural network, with an emphasis on time delay and energy reduction.
While propagating through the air, sound has a delay in its arrival, and the energy is altered mainly due to attenuation and absorption.
We introduce a novel network that models the delay and the energy reduction of the binaural speech in the Fourier space.

\subsection{Neural Fourier Shift}\label{sec:nfs}

\iffalse
If $x\in\mathcal{C}(\mathbb{R})$ integrable, we define its Fourier transform for frequency $\omega\in\mathbb{R}$ by
\begin{equation}\label{eqn:ft}
    \hat{x}(\omega) = \int_{-\infty}^\infty x(t) e^{-2\pi it\cdot\omega}\ dt.
\end{equation}
Suppose both $x$ and $\hat{x}$ are integrable, so that the inverse Fourier integral of $\hat{x}$ is well-defined.
The shift of $x$ by a value of $\delta\in\mathbb{R}$ in continuous time $x(t)\mapsto x(t+\delta)$, can be formulated in terms of the (inverse) Fourier integral:
\begin{equation}\label{eqn:continuous-shift}
    x(t+\delta) = \int_{-\infty}^\infty \hat{x}(\omega) e^{2\pi i\omega\cdot (t+\delta)}\ d\omega.
\end{equation}
The continuous shift in equation \eqref{eqn:continuous-shift} can also be formulated in a discrete version.
\fi

For any $\mathbf{x}\in\mathbb{R}^N$, define the spectrum of $\mathbf{x}$ by the Discrete Fourier Transform (DFT) $\mathbf{X}[k] := \textsc{DFT}(\mathbf{x})[k] = \sum_{m=0}^{N-1}\mathbf{x}[m]e^{-i\omega_k m}$, where $\omega_k = 2\pi k/N$ for $k=0,1,...,N-1$.
The spectrum of $\textsc{Shift}_\Delta(\mathbf{x})$, which is $\mathbf{x}$ shifted for $\Delta\in\mathbb{N}$, satisfies
\begin{equation}\label{eqn:shift-theorem}
    \textsc{DFT}\left(\textsc{Shift}_\Delta(\mathbf{x})\right)[k] = e^{-i\omega_k\Delta}\mathbf{X}[k].
\end{equation}
The equation \eqref{eqn:shift-theorem} is also well known as the (Fourier) shift theorem, which implies that a delay of $\Delta$ samples in the time corresponds to a linear phase term $e^{-j\omega_k \Delta}$ in the frequency domain.
The proposed method models the delay for early reflections by estimating the phase term.
In addition to the linear time shift where the spectral magnitude remains unaffected, we aim to model the magnitude change from the monaural to the binaural speech.

Let $\mathbf{X}\in\mathbb{C}^N$ be the spectrum of a monaural speech $\mathbf{x}$, and suppose that the source's position $p\in\mathbb{R}^3$ and the orientation $q\in\mathbb{R}^4$ are given.
We introduce a neural network $\Psi_\theta: (p,q) \mapsto (\sigma_\text{L},\phi_\text{L},\sigma_\text{R},\phi_\text{R})$ with parameter $\theta$ to estimate the magnitude change $\sigma\in\mathbb{R}^{C\times N}$ and the phase shift $\phi\in\mathbb{R}^{C\times N}$ with $C$ number of channels for left (L) and right (R) ears.
The estimated output shifts the monaural speech by $\phi$ and scales by $\sigma$ using the equation \eqref{eqn:shift-theorem}, which can be expressed as
\begin{equation}\begin{split}\label{eqn:spec}
    \hat{\mathbf{X}}_\text{L}[k] &= \sum_{c=0}^{C-1} {\sigma}_\text{L}[c,k] e^{-i\omega_k{\phi}_\text{L}[c,k]}\mathbf{X}[k],\\
    \hat{\mathbf{X}}_\text{R}[k] &= \sum_{c=0}^{C-1} {\sigma}_\text{R}[c,k] e^{-i\omega_k{\phi}_\text{R}[c,k]}\mathbf{X}[k].
\end{split}\end{equation}
% The binaural spectrum is then obtained by a concatenation as $\hat{\mathbf{X}} = \hat{\mathbf{X}}_\text{L} \oplus \hat{\mathbf{X}}_\text{R}$.
% The binaural speech $\hat{\mathbf{x}}$ is finally estimated by imposing the inverse DFT (IDFT) to the $\hat{\mathbf{X}}$.
The binaural speech $\hat{\mathbf{x}}$ is estimated by concatenation of the two $\hat{\mathbf{x}}_\text{L} \oplus \hat{\mathbf{x}}_\text{R}$ which are obtained by imposing the inverse DFT (IDFT) to $\hat{\mathbf{X}}_\text{L}$ and $\hat{\mathbf{X}}_\text{R}$.
By design, $\Psi_\theta$ estimates the frame-wise multichannel spectrum independent of the source spectrum.

For given binaural recording $\mathbf{y}$ of monaural speech $\mathbf{x}$, the neural network $\Psi_\theta$ is trained to minimize the loss between $\hat{\mathbf{x}}$ and $\mathbf{y}$ as
\begin{equation}\begin{split}
    \mathcal{L}(\hat{\mathbf{x}},{\mathbf{y}}) &= 
    \lambda_{1}\underbrace{\left\|\hat{\mathbf{x}} - \mathbf{y}\right\|_2}_{\ell_2}
    + \lambda_{2}\underbrace{\|\angle\hat{\mathbf{X}} - \angle\mathbf{Y}\|_1}_{\mathcal{L}_{\text{phs}}}\\
    &+ \lambda_{3}\underbrace{\left\|\text{IID}(\hat{\mathbf{x}}) - \text{IID}(\mathbf{y})\right\|_2}_{\mathcal{L}_{\text{IID}}}
    + \lambda_{4}\underbrace{\text{MRSTFT}(\hat{\mathbf{x}}, \mathbf{y})}_{\mathcal{L}_{\text{STFT}}},\\
\end{split}\end{equation}
where we define the IID as the mean of the interaural difference between the log magnitudes of the left and right spectrums
\begin{equation}
    \text{IID}(\mathbf{x}) :=
    \frac{1}{M}\sum_{k=0}^{M}\log_{10}\left|\mathbf{X}_{\text{L}}[k]\right|  - \log_{10}\left|\mathbf{X}_{\text{R}}[k]\right|.
\end{equation}
MRSTFT stands for the multi-resolution STFT loss \cite{yamamoto2020parallel}, and $\lambda_i$ are the hyperparameters.
$\mathcal{L}_{\text{phs}}$ denotes the angular phase error, which Richard \textit{et al.} \cite{richard2020neural} proved advantageous in overcoming the deficiencies of optimization through $\ell_2$-loss between the waveforms.

%\begin{table*}[t]
\begin{table}[t]
    \centering
    \footnotesize
    \begin{tabular}{l|ccccc}\toprule
        \textbf{Model} &
        \textbf{$\ell_2\cdot10^3$} & {\tt Amp} & $\mathcal{L}_{\text{phs}}$ & {PESQ} & $\mathcal{L}_{\text{STFT}}$
        \\\midrule
        %DSP & 1.543 & 0.097 & 1.596 & 1.610 & 2.750
        %\\
        WaveNet \cite{oord2016wavenet} & 0.179 & 0.037 & 0.968 & 2.305 & 1.915
        \\
        IR-MLP \cite{richard2022deep} & 0.236 & 0.042 & 0.933 & - & -
        \\
        WarpNet \cite{richard2020neural} & 0.167 & 0.048 & \textbf{0.807}$^*$ & - & -
        \\
        WarpNet$^\diamond$ \cite{leng2022binauralgrad} & \textbf{0.157}$^\dagger$ & 0.038 & 0.838 & \textbf{2.360}$^\dagger$ & 1.774
        \\
        {\scriptsize BinauralGrad} \cite{leng2022binauralgrad} & \textbf{0.128}$^*$ & \textbf{0.030}$^*$ & \textbf{0.837}$^\dagger$ & \textbf{2.759}$^*$ & 1.278
        \\
        \midrule
        \textbf{Ours (NFS)} & 0.172 & \textbf{0.035}$^\dagger$ & 0.999 & 1.656 & \textbf{1.241}$^*$
        \\
        \quad {\tt\footnotesize wo.NI} & 0.172 & \textbf{0.035}$^\dagger$ & 0.959 & 1.651 & 2.280
        \\
        \quad {\tt\footnotesize wo.LFF} & 0.207 & 0.037 & 1.118 & 1.592 & 1.267
        \\
        \quad {\tt\footnotesize wo.GeoWarp} & 0.197 & \textbf{0.035}$^\dagger$ & 1.200 & 1.681 & \textbf{1.248}$^\dagger$
        \\
        \quad {\tt\footnotesize wo.Shifter} & 0.433 & 0.039 & 1.570 & 1.502 & 1.288
        \\\bottomrule
    \end{tabular}
    \vspace{-3mm}
    \caption{Quantitative results.
    The highest score for each metric is marked with an asterisk ($\cdot^*$) and the second highest score is marked with a dagger ($\cdot^\dagger$).
    WarpNet$^\diamond$ shows results reported by \cite{leng2022binauralgrad}.
    }
    \vspace{-4mm}
    \label{tab:objective}
\end{table}
%\end{table*}

\subsection{Network Architecture}
As illustrated in Figure \ref{fig:system}, NFS takes three inputs: monaural audio, the source's position with orientation, and frame-wise geometric delay.
First of all, the monaural audio is chunked into frames with overlaps, where we call it as {\tt\footnotesize unfold}.
Using the position and the orientation of the source, 
NFS binauralize each frame in the Fourier space, following the formulations in Section \ref{sec:nfs}.
We illustrate the schematic network architecture of NFS in Figure \ref{fig:nfs}.

The position and the orientation of the source are encoded using sinusoidal encoding and learned Fourier features ({\tt\footnotesize LFF}) \cite{tancik2020fourier}.   
We estimate the framewise frequency response $\sigma$ and phase delay $\phi$ using the neural networks named as {\tt\footnotesize Scaler} and {\tt\footnotesize Shifter}, respectively.
The architecture of {\tt\footnotesize Scaler} and {\tt\footnotesize Shifter} is the same, which consists of cross-attention \cite{vaswani2017attention}, squeeze-and-excitation \cite{hu2018squeeze}, gMLP\cite{liu2021pay}, and channel-wise linear layers as the trainable modules.

Cross-attention fuses information of two separate embeddings using attention mechanism \cite{vaswani2017attention}.
Putting the embeddings of the source positions as query and those of the orientations as key and value, the cross-attention layer conditionally fuses the orientation to the positional information. 
The position embedding is added to the cross-attention output by a residual connection.
A channel-wise linear layer follows to project the embedding to a higher dimensional space with {\tt\footnotesize chan} number of channels (which is the same number of channels that we repeat the source during {\tt\footnotesize ch-tile}).

Squeeze-and-excitation then extracts informative features by fusing spatial and channel-wise information \cite{hu2018squeeze}.
gMLP is a variation of MLP which can be a simple alternative to the multi-head self-attention layers in Transformers \cite{vaswani2017attention,liu2021pay}.
Equipped with a spatial gating unit, gMLP captures spatial information across embeddings.
The output of gMLP is then linearly interpolated to {\tt\footnotesize freq} number of dimensions to represent the scales (for {\tt\footnotesize Scaler}) or the shifts (for {\tt\footnotesize Shifter}) for the {\tt\footnotesize freq} number of frequency bins.
The outputs of {\tt\footnotesize Scaler} and {\tt\footnotesize Shifter} are then transformed through nonlinear activations, where we denote by $\varphi:=\mathrm{Sigmoid}({\tt\footnotesize Shifter})\in\mathbb{R}^{{\tt\footnotesize chan}\times{\tt\footnotesize freq}}$ and $\varsigma:=\mathrm{Softplus}({\tt\footnotesize Scaler})\in\mathbb{R}^{{\tt\footnotesize chan}\times{\tt\footnotesize freq}}$.
The transformation is intended to enforce $\varsigma$ to be positive definite, and $\varphi$ to be in a millisecond no larger than half of the frame length.

The {\tt\footnotesize Shifter} output is biased by the geometric delay (given by a scalar $g\in\mathbb{R}$ given for each frame) computed from the geometric warp field.
We denote the biased shift by $\phi=\varphi+g$.
To enforce the energy to be inversely proportional to the power of distance, we divide the {\tt\footnotesize Scaler} output by $\phi^{2}$ (since $\phi$ is proportional to the distance), to compute the scale $\sigma=\varsigma\odot\phi^{-2}$.
Finally, we acquire the frame-wise multichannel spectrum $\sigma\exp(-i\omega\phi)\in\mathbb{C}^{{\tt\footnotesize chan}\times{\tt\footnotesize freq}}$, and multiply it to the source in the Fourier space as equation \eqref{eqn:spec}.

\newpage
After the estimated $\hat{\mathbf{X}}\in\mathbb{C}^{{\tt\footnotesize chan}\times{\tt\footnotesize freq}}$ is inverted back into the waveform using IDFT, we linearly project the channels into a single stem.
We enforce the linear projection weights to be positive definite to make the projection independent of the phase shift.
%This scaled summation of channels can be interpreted as a similar operation to approximating impulse responses.
Indeed, $\sigma\exp(-i\omega\phi)$ contains channel-wise IRs for each frame, where each channel can represent an IR for each different trajectory of acoustic rays.
The NFS output frames are then synthesized using Weighted Overlap-add (WOLA) method.
We add filtered noise to model ambient noises of the in-the-wild binaural recordings, as illustrated as {\tt\footnotesize Noise Injection} in Figure \ref{fig:nfs}.
About the effect of each network module, we elaborate in more detail based on the empirical observations in section \ref{sec:eval-quan}.

\vspace{-2mm}
\subsection{Training Details}
For each NFS responsible for left and right channels, we set both the number of channels ${\tt\footnotesize chan}$ and the number of dimensions for sinusoidal encoding to be 128.
Our model is trained using RAdam optimizer \cite{liu2019variance} for 16 epochs with a batch size of 6 where we randomly sample 800 ms of mono audio and the corresponding conditions for each batch.
We train NFS with 200 ms frame length and 100 ms hop length, and the input with 800 ms mono audio is padded and chunked into 9 frames.
While training NFS, regarding each frame as an independent batch, we stack the frames in batch dimension and estimate the output for all of the 9 frames in a single forward process.
The learning rate was initialized by $10^{-3}$ and scaled by the factor of 0.9 for every epoch.
We set $\lambda_1=10^3$, $\lambda_3=10^1$, and $\lambda_2=\lambda_4=1$.

\vspace{-2mm}
\section{Evaluation}
% \footnote{\scriptsize\url{https://github.com/facebookresearch/BinauralSpeechSynthesis/releases/tag/v1.0}}
We use the binaural speech benchmark dataset \cite{richard2020neural} for all experiments, which contains monaural-binaural paired audio data for approximately 2 hours long, with 48 kHz sampling rate.
The binaural audio that the dataset provides is recorded using a KEMAR mannequin in a regular room, for the source speeches spoken by eight different subjects.
The position and orientation of the source are tracked at 120 Hz and are aligned with the audio.
%We follow the same validation and test split of the official benchmark.

% We use WarpNet\footnote{\scriptsize\url{https://github.com/facebookresearch/BinauralSpeechSynthesis/releases/tag/v1.1}} \cite{richard2020neural} and BinauralGrad\footnote{\scriptsize\url{https://github.com/microsoft/NeuralSpeech/tree/master/BinauralGrad\#pretrained-models}} \cite{leng2022binauralgrad} as the baselines of the comparison to evaluate our model.
We use WarpNet \cite{richard2020neural} and BinauralGrad \cite{leng2022binauralgrad} as the baselines of the comparison.
% The work by Huang \textit{et al.} \cite{huang2022end} was not included in the comparison because of the different training data.
We compare our model with the 6-layer IR-MLP \cite{richard2022deep} model with 512 hidden units they used for their experiment in training with the benchmark dataset.
%Ablation studies using our model without the time shift (denoted by {\tt\footnotesize wo.Shifter}), or without the noise injection (denoted by {\tt\footnotesize wo.NI}) are also conducted.
We also evaluate four variants of NFS: (1) NFS trained with {\tt\footnotesize Noise Injection} but inference without it denoted by {\tt\footnotesize wo.NI}, (2) our network trained without learned Fourier features denoted by {\tt\footnotesize wo.LFF}, (3) our network trained without the time shift denoted by {\tt\footnotesize wo.Shifter}, and (4) our network trained without Geometric Warpfield denoted by {\tt\footnotesize wo.GeoWarp}.

\vspace{-2mm}
\subsection{Quantitative Evaluation}\label{sec:eval-quan}

Table \ref{tab:objective} compares the objective results of each model for the binaural speech synthesis benchmark test set.
{\tt\footnotesize Amp} denotes the (spectral) amplitude error \cite{richard2020neural}.
NFS records the best MRSTFT score ($\mathcal{L}_{\text{STFT}}$) and the second-best amplitude error compared to baselines.
% Without Noise Injection
\textbf{Absence of noise injection} degrades the MRSTFT score of the NFS.
As the {\tt\footnotesize NI} module is intended to model ambient noise, inference results of NFS without it (denoted as {\tt\footnotesize wo.NI}) show higher $\mathcal{L}_{\text{STFT}}$ than the original NFS.
However, for all other metrics, the model {\tt\footnotesize wo.NI} also records a similar performance to the original NFS -- which implies that NFS learns to scale and shift the source independently of the ambient noise.
% Without LFF
\textbf{Learned Fourier features} slightly advantages NFS to encode the conditions, since {\tt\footnotesize wo.LFF} performs worse than NFS in every measure.
However, we emphasize that the performance of {\tt\footnotesize wo.LFF} for $\mathcal{L}_{\text{STFT}}$ is still better than that of BinauralGrad.
% Without Shifter
\textbf{Without Shifter}, our network performs worse than the original NFS.
{\tt\footnotesize wo.Shifter} can only adjust the frequency response for each channel where all phases are uniformly shifted by the geometric delay.
The absence of {\tt\footnotesize Shifter} degrades especially for the $\ell_2$-loss and the angular phase error, and this shows the impact of the phase shift by {\tt\footnotesize Shifter}.
% Without GeoWarp
\textbf{Without geometric delay}, NFS predicts both the early reflections and the direct arrivals.
Quantitative results show that {\tt\footnotesize wo.GeoWarp} also performs similarly to NFS, but is slightly worse in $\mathcal{L}_{\text{phs}}$.
As it appears in Figure \ref{fig:scale-vs-delay}, the relation between the input positions and the most dominant response estimated from {\tt\footnotesize wo.GeoWarp} shows similar dynamics to that of NFS.
It turns out that biasing $\varphi$ with $g$ of {\tt\footnotesize GeoWarp} advantages NFS to estimate more accurate phase shifts and magnitude scales.

%\begin{table*}[t]
\begin{table}
    \centering
    \footnotesize
    \begin{tabular}{l|cc}\toprule
        \textbf{Model} & \textbf{\# Param.} &
        \textbf{MACs} ($\downarrow$)
        \\\midrule
        BinauralGrad (stage 1, single step) \cite{leng2022binauralgrad} & 6.91 M & 229.4 G
        \\
        BinauralGrad (stage 2, single step) \cite{leng2022binauralgrad} & 6.91 M & 229.3 G
        \\
        WarpNet \cite{richard2020neural} & 8.59 M &  19.15 G
        \\
        IR-MLP \cite{richard2022deep} & 1.62 M & -
        \\
        \midrule
        \textbf{Ours (NFS)} & \textbf{0.55 M} &  \textbf{3.400 G}
        \\
        %\quad {\tt\footnotesize wo.NI} & \textbf{0.55 M} & \textbf{3.390 G}
        %\\
        %\quad {\tt\footnotesize wo.LFF} & \textbf{0.55 M} & \textbf{3.390 G}
        %\\
        %\quad {\tt\footnotesize wo.GeoWarp} & \textbf{0.55 M} & \textbf{3.400 G}
        %\\
        \quad {\tt\footnotesize wo.Shifter} & \textbf{0.28 M} & \textbf{1.700 G}
        \\
        \bottomrule
    \end{tabular}
    \vspace{-2mm}
    \caption{Comparison of memory and computational cost.
    }
    \label{tab:capacity}\vspace{-2mm}
\end{table}
%\end{table*}

\iffalse
\begin{table}
    \centering
    %\setlength{\abovecaptionskip}{5pt}
    \scriptsize
    \begin{tabular}{l|c|ccccc}\toprule
        \multirowcell{2}[-0.8ex]{\textbf{Model}} & \multirowcell{2}[-0.8ex]{\textbf{MUSHRA}} & \multicolumn{5}{c}{\textbf{$p$-value}} \\\cmidrule{3-7}%\cline{3-7}
         & & WN & BG & NFS & {\tt\scriptsize wo.NI} & {\tt\scriptsize wo.Sh}
        \\\midrule
        Ref. & $0.92_{\pm0.08}$ &
        0.0000 & 0.0000 & 0.0000 & 0.0000 & 0.0000
        \\\midrule
        WN \cite{richard2020neural} & $0.64_{\pm0.20}$ &
        -- & \textbf{1.0000} & \textbf{1.0000} & 0.0790 & 0.0005 
        \\
        BG \cite{leng2022binauralgrad} & $0.63_{\pm0.20}$ &
        \textbf{1.0000} & -- & \textbf{1.0000} & 0.2497 & 0.0040 
        \\\midrule
        NFS  & $0.61_{\pm0.20}$ &
        \textbf{1.0000} & \textbf{1.0000} & -- & 0.7509 & 0.0049 
        \\
        {\tt\scriptsize wo.NI}  & $0.58_{\pm0.18}$ &
        0.0790 & 0.2497 & 0.7509 & -- & 0.4596 
        \\
        {\tt\scriptsize wo.Sh}  & $0.52_{\pm0.21}$ &
        0.0005 & 0.0040 & 0.0049 & 0.4596 & -- 
        \\\bottomrule
    \end{tabular}
    \vspace{-2mm}
    \caption{{\footnotesize Ref.}, {\footnotesize WN}, {\footnotesize BG}, and {\tt\footnotesize wo.Sh} stands for the binaural recorded data, Warpnet, BinauralGrad, and {\tt\footnotesize wo.Shifter}, respectively.
    Post-hoc analysis of the MUSHRA test result shows no significant differences among the three models: WarpNet, BinauralGrad, and NFS.
    }
    \label{tab:mushra}\vspace{-3mm}
\end{table}
\fi

\begin{table}
    \centering
    \footnotesize
    \begin{tabular}{l|c|ccccc}\toprule
        \multirowcell{2}[-0.8ex]{\textbf{Model}} & \multirowcell{2}[-0.8ex]{\textbf{Score}} & \multicolumn{5}{c}{\textbf{$p$-value}} \\\cmidrule{3-7}%\cline{3-7}
         & & WN & BG & NFS & {\tt\footnotesize wo.NI} & {\tt\footnotesize wo.Sh}
        \\\midrule
        Ref. & $0.92_{\pm0.08}$ &
        0.000 & 0.000 & 0.000 & 0.000 & 0.000
        \\\midrule
        WN  & $0.64_{\pm0.20}$ &
        -- & \textbf{1.000} & \textbf{1.000} & 0.079 & 0.001 
        \\
        BG & $0.63_{\pm0.20}$ &
        \textbf{1.000} & -- & \textbf{1.000} & 0.249 & 0.004
        \\\midrule
        NFS  & $0.61_{\pm0.20}$ &
        \textbf{1.000} & \textbf{1.000} & -- & 0.751 & 0.005 
        \\
        {\tt\footnotesize wo.NI}  & $0.58_{\pm0.18}$ &
        0.079 & 0.249 & 0.751 & -- & 0.460
        \\
        {\tt\footnotesize wo.Sh}  & $0.52_{\pm0.21}$ &
        0.001 & 0.004 & 0.005 & 0.460 & -- 
        \\\bottomrule
    \end{tabular}
    \vspace{-2mm}
    \caption{
        Analysis of the MUSHRA score shows no significant differences among the three models: WarpNet, BinauralGrad, and NFS.
        {\footnotesize Ref.}, {\footnotesize WN}, {\footnotesize BG}, and {\tt\footnotesize wo.Sh} stands for the binaural recorded data, Warpnet, BinauralGrad, and {\tt\footnotesize wo.Shifter}, respectively.    
    }
    \label{tab:mushra}\vspace{-3mm}
\end{table}

\vspace{-2mm}
\subsection{Perceptual Evaluation}

To measure the perceptual difference of the rendered binaural speeches, we conduct a listening test using multiple stimuli with hidden reference and anchor (MUSHRA).
%\footnote{\scriptsize\url{https://github.com/audiolabs/webMUSHRA}}
We present nine questions to every nine subjects familiar with the concept of spatial audio.
Each question includes the binaural recordings of the benchmark test set as a reference.
% We instruct participants to rate how similarly spatialized the samples are, compared to the reference sample in a scale from 0 to 1.
The participants are asked to rate how similarly spatialized the samples are, compared to the reference sample on a scale from 0 to 1.
The samples include hidden references as a high-anchor, and the outputs of the models appear in random permutation order.
We put {\tt\footnotesize wo.Shifter} as a low-anchor with a structure that can only change the frequency response.

% Figure \ref{fig:mushra} shows the MUSHRA scores.
%Score differences between NFS and other models may appear insignificant in the figure, but analysis using multiple posthoc paired t-tests with Bonferroni corrections reveals differences between models.
Table \ref{tab:mushra} shows the result of the post hoc analysis with Bonferroni correction for the multiple paired t-test.
We find a significant difference with $p<0.005$ between {\tt\footnotesize wo.Shifter} and others including NFS, WarpNet, and BinauralGrad.
This again highlights the importance of {\tt\footnotesize Shifter} in NFS.
{\tt\footnotesize wo.NI} shows a poor $\mathcal{L}_{\text{STFT}}$ in Table \ref{tab:objective}, but according to our posthoc analysis, it is found to have a perceptually insignificant difference between NFS, WarpNet, or BinauralGrad.
Although NFS scored worse PESQ in quantitative evaluation, the difference between NFS, WarpNet, and BinauralGrad is found insignificant from our analysis of MUSHRA scores.
We conclude that NFS's capability in modeling ambient noise can be limited, but it is trivial when it comes to spatialization perception.

\iffalse
\begin{figure}[b]
    %\vspace{-7.3mm}
    \vspace{-8mm}
    \begin{minipage}[b]{1.0\linewidth}
        \centering
        \centerline{\includegraphics[width=0.75\linewidth]{img/mushra.pdf}}
    \end{minipage}
    \vspace{-6.8mm}
    \caption{Listening test violin plots.
    {\tt\footnotesize B.Grad} denotes BinauralGrad.
    }
    \vspace{-5mm}
    \label{fig:mushra}
\end{figure}
\fi

\vspace{-2mm}
\subsection{Efficiency and Interpretability}
Besides recording decent performance in quantitative and qualitative evaluations, NFS is highly efficient in memory and computational cost, which is a huge advantage for commodity hardware.
% Table \ref{tab:capacity} compares the number of parameters and multiply-accumulates (MACs)\footnote{\scriptsize\url{https://github.com/sovrasov/flops-counter.pytorch}}, which is a widely used metric to measure the number of computations of a model.
Table \ref{tab:capacity} compares the number of parameters and multiply-accumulates (MACs).
% MAC is a widely used metric to measure the number of computations of a model.
Among the comparative models, NFS is 25 times lighter than BinauralGrad, which consists of a total of 13.8 million parameters.
Compared to IR-MLP, our method is lighter and performs better in most objective evaluations.
% NFS also shows the lightest computational cost.
% NFS requires only 17.8\% of the computation compared to WarpNet, and 1.48\% compared to a single denoising step in BinauralGrad (stage 1).
The number of computations required for NFS reaches only 17.8\% of that of WarpNet, which is only 1.48\% of that of a single denoising step in BinauralGrad (stage1).
% Note that BinauralGrad inference through multiple denoising steps for each stage, by its design.
% We do not compare the MACs of IR-MLP as there is no publicly available model.

\begin{figure}[t]
    \begin{minipage}[b]{0.5\linewidth}
        \centering
        \centerline{\includegraphics[width=1.08\linewidth]{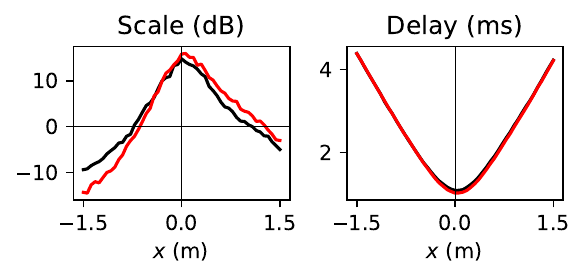}}
        \vspace{-2mm}
        \centerline{(a) Longitudinal (NFS)}
    \end{minipage}
    \begin{minipage}[b]{0.5\linewidth}
        \centering
        \centerline{\includegraphics[width=1.08\linewidth]{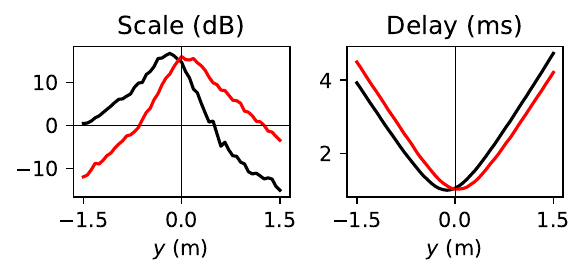}}
        \vspace{-2mm}
        \centerline{(b) Lateral (NFS)}
    \end{minipage}
    \begin{minipage}[b]{0.5\linewidth}
        \centering
        \centerline{\includegraphics[width=1.08\linewidth]{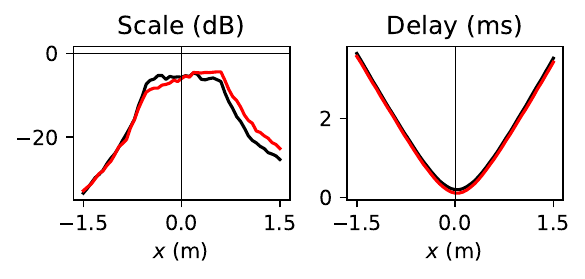}}
        \vspace{-2mm}
        \centerline{(c) Longitudinal ({\tt\footnotesize wo.GeoWarp})}
    \end{minipage}
    \begin{minipage}[b]{0.5\linewidth}
        \centering
        \centerline{\includegraphics[width=1.08\linewidth]{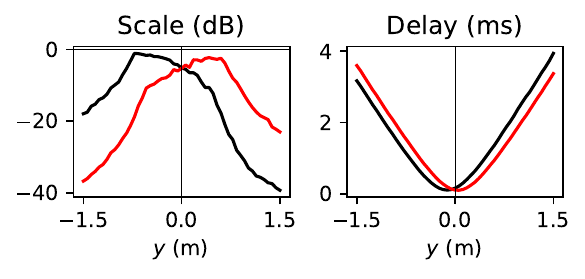}}
        \vspace{-2mm}
        \centerline{(d) Lateral ({\tt\footnotesize wo.GeoWarp})}
    \end{minipage}
    \vspace{-5mm}
    \caption{Plot of $\sigma$ and $\phi$ averaged over the covering frequencies for the channel with the most dominant intensity.
    Responses for the left and right channel is colored in black (---) and red ({\color{red} ---}), respectively.
    }
    \vspace{-5mm}
    \label{fig:scale-vs-delay}
\end{figure}

Estimation within the Fourier space also makes our model easier to interpret.
Figure \ref{fig:scale-vs-delay} shows the estimated output for the positions along the trajectories parallel to the longitudinal and lateral axes.
The delay from NFS is not very different from the geometric delay which is proportional to the Euclidean distance between the source and the receiver.
This is natural in that the direct arrivals from the source are the most dominant in their intensity.
The scale decreases sharply at locations where the direct path is obscured by the pinna or head.
Even without the guidance of {\tt\footnotesize GeoWarp}, our network reflects similar physics in its estimation.
This interpretability directly demonstrates how precisely our deep neural network approximates the environment we are trying to simulate and allows us to further compensate for the weaknesses of the model.

\iffalse
\input{tab/ood-subjective}
\subsection{Out-of-distribution Robustness}
We also perceptually evaluate models using out-of-distribution (OOD) data including singing, music, and general noise.
We mixdown the OOD data to mono with RMS normalization, and render them using the positions given in the test set of the binaural speech benchmark dataset.
We sample `music' and `noise' subsets from MUSAN \cite{musan2015} and sample `vibrato' subset from VocalSet \cite{wilkins2018vocalset} as the evaluation for the singing data.
Table \ref{tab:ood} compares mean opinion score (MOS) for the estimated outputs.
\fi

\vspace{-2mm}
\section{Conclusion}
\vspace{-1mm}
We present NFS, a lightweight neural network for binaural speech rendering.
Building upon the foundations of the geometric time delay, NFS predicts frame-wise frequency response and phase lags for multiple early reflection paths in the Fourier space.
Defined in the Fourier domain, NFS is highly efficient and operates independently of the source domain by its design.
Experimental results show that NFS performs comparable to the previous studies on the benchmark dataset, even with its lighter memory and fewer computations.
NFS is interpretable in that it explicitly displays the frequency response and phase delays of each acoustic path, for each source position.
We expect improving NFS using generic binaural audio datasets to generalize to arbitrary domains as our future works.
%We also experiment with NFS for binauralizing the out-of-distribution dataset including singing, music, and general noise data.

\vspace{-2mm}
\section{Acknowledgement}
\vspace{-1mm}
This work was supported by Institute of Information \& Communications Technology Planning \& Evaluation (IITP) grant funded by the Korean Government 2022-0-00641.

\vfill\pagebreak
\newpage

\bibliographystyle{IEEEbib}
\bibliography{refs}

\end{document}